\documentclass[11pt]{article}

\usepackage{amsxtra,amssymb,amsthm,amsmath,latexsym}
\usepackage{graphicx}
\usepackage{epsfig}
\usepackage{afterpage}

\newcommand{\nb}{\nabla}

\newcommand{\bee}{\begin{equation*}}
\newcommand{\eee}{\end{equation*}}
\newcommand{\be}{\begin{equation}}
\newcommand{\ee}{\end{equation}}

\title{Permeabilities of metamaterials}
\author{A.G. Ramm \\
\small Department of Mathematics\\[-0.8ex]
\small Kansas State University, Manhattan, KS 66506-2602, USA\\
\small \texttt{ramm@math.ksu.edu}}

\date{}
\begin{document}

\maketitle

\begin{abstract}
Scattering of electromagnetic (EM) waves by many small particles,
embedded in a given medium, is studied. Physical properties of
the particles are described by their boundary impedances. The
limiting equation is  obtained
for the effective EM field in the limiting medium, 
in the limit $a\to 0$, where $a$ is the characteristic size of a
particle and the number $M(a)$ of the particles tends to infinity at
a suitable rate. 
An analytical formula for the permeability 
$\mu(x)$ of the limiting medium is given. 
Analysis of this formula allows one to find out the range of the values of 
the permeability in the material, obtained
by embedding many small particles.

\end{abstract}
 



\section{Introduction}
In this paper we outline a theory of monochromatic electromagnetic (EM) 
wave scattering by many small  particles (bodies) embedded in a 
homogeneous medium which is described by the permittivity 
$\epsilon'=\epsilon_0+i\frac \sigma \omega$, $\epsilon_0>0$,
and constant permeability $\mu_0$. The wave number $k=\omega (\epsilon' 
\mu_0)^{1/2}$.  Related papers are \cite{R581}-\cite{R598},
and the results we use without proof are taken mostly from \cite{R581}.
The small particles are embedded in a finite domain $\Omega$.
Smallness of the particles means that $|k|a\ll 1$, where $a$ is the 
characteristic size of the particles, and $k$ is the wave number. 
The medium, created by  the embedding of the small particles, has new 
physical properties. 
In particular, it has a spatially inhomogeneous magnetic permeability 
$\mu(x)$, which can be controlled by the choice of
the boundary impedances of the embedded small particles and their
distribution density.  This is 
a new physical effect.
An analytic formula for the permeability of the new medium is derived
in \cite{R581}:
\be\label{e1}\mu(x)=\frac {\mu_0}{\Psi(x)},
\ee
where
\be\label{e12}\Psi(x)=1+\frac{8\pi i}{3 \mu_0 \omega} h(x) 
N(x).\ee 
Here $\omega$ is the frequency of the EM field, $\epsilon_0$
is the constant dielectric parameter of the original medium, 
$h(x)$ is a function describing boundary impedances of the small
embedded particles, and $N(x)\geq 0$ is a function describing the 
distribution of these particles.  
We assume that in any subdomain $\Delta$, the number
$\mathcal{N}(\Delta)$ of the embedded particles $D_m$ is given by the
formula: 
\be\label{e3}
\mathcal{N}(\Delta)=\frac{1}{a^{2-\kappa}}\int_{\Delta} 
N(x)dx[1+o(1)],\quad
a\to 0, \ee 
where $N(x)\geq 0$ is a continuous function, vanishing
outside of the finite domain $\Omega$ in which
small particles (bodies)  $D_m$ are distributed, $1\leq m \leq M$,
$M=M(a)$, 
$\kappa\in(0,1)$ is a parameter, and
the boundary impedances of the small particles are defined by the
formula
\be\label{e4}\zeta_m=\frac{h(x_m)}{a^\kappa},\quad x_m\in D_m,\ee
where $x_m$ is a point inside $m-$th particle $D_m$, Re $h(x)\geq 0$, and 
$h(x)$ is a continuous function vanishing 
outside $\Omega.$ The impedance boundary condition on the surface $S_m$
of the $m-$th particle $D_m$ is $E^t=\zeta_m [H^t,N]$, where $E^t$ ($H^t$)
is the tangential component of $E$ ($H$) on $S_m$, and $N$ is the unit
normal to $S_m$, pointing out of $D_m$. Physical properties of the 
impedance $\zeta$ are discussed in \cite{LL}. In particular,
Re$\zeta\ge 0$.  
 
Since one can choose the functions $N(x)$ and $h(x)$, one can create a 
desired magnetic permeability in $\Omega$. This is a novel idea, to the 
author's knowledge.

We have also derived in \cite{R581} an analytic 
formula for the refraction coefficient of the medium in $\Omega$
created by the embedding of many small particles.
\be\label{e5}
K^2(x)=\frac{k^2}{1+\frac{8\pi i}{3 \mu_0 \omega}  h(x)N(x)},\quad
k^2=\omega^2 \epsilon' \mu_0, \ee
where the coefficient $\frac{16\pi}{3}$ appears if $D_m$ are balls of 
radius $a$
centered at the points $x_m$. For the small bodies $D_m$ of
arbitrary shape the coefficient $\frac{8\pi}{3}$ should be replaced by 
a tensorial coefficient 
$c_{m}=O(1)$, depending of the shape of $D_m$.
  
An equation for the EM field in the limiting medium is derived in 
\cite{R581}:
\be\label{e6}
E(x)=E_0(x)-\frac{8\pi i}{3 \mu_0 \omega} 
 \nb\times \int_{\Omega}g(x,y)h(y)N(y)\nb\times
E(y)dy. \ee
The limiting medium is created  when the size $a$ of small particles tends 
to zero 
while the total number $M=M(a)$ of the particles tends to infinity at the 
rate determined by \eqref{e3}.

The refraction coefficient in the limiting medium is spatially 
inhomogeneous.

Our theory may be viewed as a "homogenization theory", but it differs
from the usual homogenization theory (see, e.g., \cite{CD}, \cite{MK},
and references therein) in several respects: we do not assume any periodic 
structure in the distribution of small bodies, our operators are 
non-selfadjoint, the spectrum of these operators is not discrete,
etc. Our ideas, methods, and techiques are quite different from the usual 
methods. 

These ideas are similar to  the ideas developed in
papers \cite{R509, R536}, where scalar wave scattering by small
bodies was studied, and in the papers \cite{R581} and \cite{R598},
where EM wave scattering by many small particles was stidied.
Scattering of EM waves brought new technical difficulties
which were resolved in \cite{R581} and \cite{R598}. The difficulties come 
from the vectorial 
nature of the boundary conditions. Results  from 
\cite{R563} were used for a justification of the passages to the limit
$a\to 0$. 

In \cite{R581} a new numerical method for solving many-body 
wave-scattering
problems for small scatterers is given.

We predict theoretically the new physical phenomenon:
by embedding many small particles with suitable boundary impedances 
into a given homogeneous medium, one can create a medium with a desired
spatially inhomogeneous permeability \eqref{e1}.

One can create material with a desired permeability $\mu(x)$
by embedding small particles with suitably chosen boundary impedances.
Indeed, by formula \eqref{e1} one can choose a complex-valued, in 
general, function $h(x)$, and a non-negative function  $N(x)\ge 0$,
describing the density distribution of the small particles,
so that the right-hand side of formula \eqref{e1} will yield a desired 
function $\mu(x)$.  

The goal of this paper is to discuss the possible values of magnetic
permeability one can get in the metamaterial, obtained by embedding many 
small impedance particles in a given material.

This discussion is given in the next Section. 

\section{ Possible values of permeability}

Let us denote $q=q(x):=\frac{8\pi N(x)}{3\omega \mu_0}$, and  
rewrite formula \eqref{e1} as
\be\label{e7}
\mu(x)=\frac {\mu_0( 1-qh_2 -iqh_1)} {( 1-qh_2)^2+q^2h_1^2} 
\ee
Quantity $q(x)>0$. Thus, the values of $\mu(x)$ can be controlled by 
choosing the function $h$ in equation \eqref{e4}. 

{\bf Conclusion:}

{\it Assume that the impedances of the small particles are defined  
by formula \eqref{e4}, where $h(x)=h_1(x)+ih_2(x)$ is a function with
Re$h(x)=h_1(x)\geq 0$. Let $\mu(x)=\mu_1(x)+i\mu_2(x)$, where 
$\mu_1(x)$
is the real part of $\mu(x)$. Then
it is not possible to create metamaterial with $\mu_2(x)>0$
by embedding many small impedance particles into a given material. 
Indeed, 
$$\mu_2(x)= -\frac{\mu_0qh_1(x)} {( 1-qh_2(x))^2+q^2h_1^2(x)}<0,$$
provided that $\mu_0$, $q$ and $h_1$ are all positive, which is the case. 
The assumption $h_1(x)\geq 0$ is a necessary assumption for
boundary impedances, see \cite{LL}, p.301. 

For high frequencies $\omega$ the quantity $q(x)=O(\frac 1 {\omega})$
is small, so if $h_1(x)=O(1)$, then $\mu_2(x)=O(\frac 1 {\omega})$
is also small.
}

\end{document}